# Photoionization of Endohedrals with account of fullerenes polarization

**M. Ya. Amusia[1, 2][1] and L. V. Chernysheva[2]**

[1]*Racah Institute of Physics, the Hebrew University, Jerusalem 91904, Israel*
[2]*A. F. Ioffe Physical-Technical Institute, St. Petersburg 194021, Russian Federation*

**Abstract:**

We have calculated photoionization cross-section of endohedral atoms $A@C_N$. We took into account the polarizability of the fullerene electron shell $C_N$ that modifies the incoming photon beam and the one-electron wave functions of the caged atom A. We employ simplified versions of both static and dynamic polarization.

The properly modified one-electron wave functions became a starting point of the account of the multi-electron correlations in the frame of the random phase approximation with exchange (RPAE). We treat atomic and fullerenes polarization that act upon photoelectron similarly, substituting them by static polarization potential. The photon beam polarizes the fullerene. This effect is accounted for introducing a polarization factor.

As concrete objects, we have considered Ar and Xe atoms inside fullerene $C_{60}$. Inclusion of polarization prominently increases the photoionization cross-section mainly close to the threshold.

**1**. We demonstrate in this Letter that the effect of polarizability of the fullerenes $C_N$ electron shell strongly affects the caged atom A photoionization cross-section not only via the modification of the incoming photon beam that acts upon atom A located inside $C_N$. The modification of the photon beam was qualitatively understood and quantitatively described relatively long ago (see [1] and references therein). It is expressed via big (compared to atomic values) dynamic dipole polarizability $\alpha_d(\omega)$ of $C_N$, where $\omega$ is the photon energy.

However, the static polarizability of the fullerene $C_N$ is also big, and is able to strongly modify the electron wave functions of the atom A inside $C_N$ and $C_N$ itself, thus leading to prominent new features in photoionization cross-section of $A@C_N$. Clarification of this effect is the principal aim of this Letter.

The investigation performed here became possible as a result of recent prominent success in studies of the electron scattering upon $C_N$ and $A@C_N$ [2, 3]. These papers present relatively simple methods that permitted to evaluate the role of $C_N$ polarizability upon the cross-section of elastic scattering of an electron by $A@C_N$. The idea is to take into account the polarization potential of both $C_N$ and A, and to clarify their influence upon photoionization cross-section. At this moment we are unable to make one, but quite a big step further and to take into account the role of interaction between electrons that belong to A and $C_N$. [2].

[1]amusia@vms.huji.ac.il

[2] We employ here the atomic system of units $m = e = \hbar = 1$. Here $m$ is the electron mass; $e$ is its charge and $\hbar$ is the Planck constant.

**2.** We intend to calculate photoionization cross-sections of A@$C_N$ at first in the frame of one-electron Hartree-Fock (HF) approximation, and then add interelectron correlations in the frame of the random phase approximation with exchange (RPAE). The necessary details about HF and RPAE equation and their solutions one can find in [4]. Here we present only important definitions and the main points of calculation procedures.

The endohedral A@$C_N$ we treat as an atom A located inside of a polarizable shell $C_N$, the direct electronic structure of which we neglect. We represent $C_N$ by a static potential with minimal number of adjustable parameters. Very often it is considered as a square well, with thickness $\Delta$, depth, $-U_o$ and inner (outer) radiuses $R \mp \Delta/2$ [4], with $R$ being the $C_N$ radius (see, e.g. [5]).

Such a potential leads, however, to unrealistic shape of the charge distribution [6].-This is why in this Letter we choose another, so-called Lorentz-shape fullerene potential that is suggested in [6]:

$$U_C(r) = -U_o R \frac{d}{(r-R)^2 + d^2}. \quad (1)$$

Here $d$ is the potential width.

Along with $U(r)$, electrons of atom A are under the action of $C_N$ polarization potential $W$. In general, this potential is non-local, dependent upon scattering energy $E$, and complex, i.e. has along with a real, also an imaginary part. To derive such a potential from first principles is very difficult, if possible at all. However, it is much easier to use instead a semi-empirical polarization potential $W_C$.

A simplified approach that is often employed in studies of electron-atom collisions (see, e.g. [4]), suggests the following simplified version of $W_C \equiv W_C(r)$:

$$W_C(r) = -\frac{\alpha_d^C}{(r^2 + b_C^2)^2}. \quad (2)$$

Here $\alpha_d^C$ is the static dipole polarizability of $C_N$, and $b_C$ is a cut-off parameter $b_C \approx R$.

This version was applied to electron-endohedral scattering [2, 3]. It reproduces quite well the asymptotic behavior of $W(r) \sim -1/r^4$ at $r \to \infty$. The important contribution to scattering phases comes from big distances. This is why for electron-fullerene scattering it is inessential that at $r \le b_C$ $W_C(r)$ noticeably increases by the absolute value, from $\alpha_d^C / 4b_C^4$ at $r = b_C$ to $\alpha_d^C / b_C^4$ at $r = 0$.

But $W_C$ cannot increase in the empty space inside a fullerene. In photoionization of an endohedral, the region inside the fullerene is important since the following inequalities $r_a < r < R$ are valid, where $r_a$ is the radius of the atom A. For sufficiently big fullerenes, such as $C_{60}$, with N=60, and bigger, and for most of the atoms in their ground states, stronger inequalities $r_a \ll r \ll R$ are valid. Therefore, in photoionization studies we recommend to employ instead of (2) the following formula:

$$W_{C(s)}(r)\Big|_{r>b_C} = -\frac{\alpha_d^C}{(r^2+b_C^2)^2} \text{ and } W_{C(s)}(r)\Big|_{r<b_C} = -\frac{\alpha_d^C}{4b_C^4}. \quad (3)$$

Since we intend to include $W_C(r)$ in calculation of endohedral photoionization, in place is to include also the possible contribution of polarization potential of the atom A, $W_A(r)$. This could be achieved substituting in Eq. (2) $\alpha_d^C$ and $b_C$ by respective atomic values - $\alpha_d^A$ and $b_A$ that leads to the following expression for $W_A(r)$:

$$W_A(r) = -\frac{\alpha_d^A}{(r^2+b_A^2)^2}. \quad (4)$$

This expression should not be corrected at small distances (compare (2) and (3)), since there the atomic HF potential is much bigger than the polarization one, contrary to the fullerene case.

In electron-atom scattering process much more accurate than employment of Eq. (4) is the approach, based on the many-body perturbation theory [4]. However, this approach was never (except literally in a couple of papers, [7, 8]) used in studies of atomic photoionization, although it permits to take into account on equal ground not only the polarization action of atomic electrons upon the outgoing one, but also the modification of the photon-electron interaction by accounting excitations of core electrons. The explanation of rare use of this approach is in its complexity even for atoms. To do the same for endohedrals is much more complicated. This is why we decided in studies of photoionization of endohedrals to treat the polarization of fullerene $C_N$ and atom A on equal ground, using Ed. (3) and (4).

As a result of presented approximations, the HF equation for multi-electron atoms looks like

$$\begin{aligned}&\left[-\frac{\Delta}{2}-\frac{Z}{r}+U_C(r)+W_C(r)+W_A(r)\right]\phi_j(x)\\&+\sum_{k=1}^{N_A}\int\phi_k^*(x')\frac{dx'}{|\mathbf{r}-\mathbf{r}'|}\left[\phi_k(x')\phi_j(x)-\phi_j(x')\phi_k(x)\right]=E_j\phi_j(x)\end{aligned}. \quad (5)$$

Here Z is the nuclear charge, $\phi_j(x)$ is the one-electron wave function, $x \equiv \vec{r},\vec{\sigma}$ are the combination of electron coordinate and spin variables, $E_j$ is the one-electron or so-called HF energy; the summation is performed over all occupied electron states $N_A$ of the atom A.

The photoionization cross-section of A@$C_N$ in HF $\sigma_i^{r,\nabla}(\omega)$ is determined by the square module of dipole matrix elements in the length $\vec{\varepsilon}\vec{r}$ or velocity $\vec{\varepsilon}\vec{\nabla}$ forms, calculated between HF wave functions (5) of the electron that undergoes transition from the initial state *i* to the final continuous spectrum state *k* due to photon absorption [4]:

$$d_{ik}^{r} = \omega\int \phi_i^*(x)(\vec{\varepsilon}\vec{r})\phi_k(x)dx,\;\; d_{ik}^{\nabla} = \int \phi_i^*(x)(\vec{\varepsilon}\vec{\nabla})\phi_k(x)dx,\;\; \omega \equiv E + I_i\,. \tag{6}$$

Here $E$ is the energy of states $k$ and $I_i$ is the ionization potential of the subshell $i$.

The following expression determines the photoionization cross-section $\sigma_i^{r,\nabla}(\omega)$ of a one-electron transition $i \to k$:

$$\sigma_i^{r,\nabla}(\omega) = \frac{4\pi^2}{c\omega}\int \left|d_{ik}^{r,\nabla}\right|^2 \delta\left(\omega - I_i - E\right)dE,\;. \tag{6}$$

Relations similar to (5) and (6) give the photoionization cross-section in RPAE, if one substitutes the HF matrix elements $d_{if}^{r,\nabla}$ by solutions of RPAE equations

$$\left\langle i\left|D(\omega)\right|k\right\rangle = \left\langle i\left|d^{r,\nabla}\right|k\right\rangle + \left(\sum_{\nu\le F,\nu'>F} - \sum_{\nu>F,\nu'\le F}\right)\frac{\left\langle \nu'\left|D(\omega)\right|\nu\right\rangle\left\langle \nu i\left|V\right|\nu' k - k\nu'\right\rangle}{\left[\omega - E_{\nu'} + E_{\nu} \pm i\delta\right)\right]}. \tag{7}$$

Here $V$ denotes the Coulomb inter-electron interaction, sums over $\nu \le F$ includes occupied one-electron states, while sums over $\nu > F$ includes excited discrete levels and integration over continuous excitation energies of the atom A in A@$C_N$. In the denominator the sign $\pm$ means + for $\nu'$ vacant and – for $\nu'$ occupied one-electron states, respectively. Note that $D$ does not have $r,\nabla$ indexes, since in RPAE corresponding values are equal [4].

Till now we have neglected the corrections to RPAE due to polarization of the fullerene by the incoming photon beam. This can be done assuming that a strong inequality between atomic $r_a$ and fullerene $R$ radiuses $r_a \ll R$ holds. The small parameter $r_a / R \ll 1$ permits to present corrections to the RPAE amplitude $\left\langle i\left|D(\omega)\right|k\right\rangle \equiv D_{ik}(\omega)$ due to fullerene polarization as a series in $r_a / R \ll 1$, retaining the lowest term only. As a result, the photoionization amplitude of an endohedral $D_{ik}^{A@C_N}(\omega)$ can be presented as a product of amplitude (7) and an only $\omega$ dependent factor $G^C(\omega)$ [1]:

$$D_{ik}^{A@C_N}(\omega) = G^C(\omega)D_{ik}(\omega) \equiv \left[1 - \frac{\alpha_d^C(\omega)}{R^3}\right]D_{ik}(\omega)\,. \tag{8}$$

Here $\alpha_d^C(\omega)$ is the dynamic dipole polarizability of the $C_N$.

**3.** We performed calculations using computing codes of the system ATOM-M [9]. As concrete objects of investigation, we choose endohedrals Ar@$C_{60}$ and Xe@$C_{60}$. Fig. 1 presents the potentials $U_C(r)$, $W_C(r)$, $W_{C(s)}(r)$ and $W_A(r)$. Fig. 2 - 5 depict the results of our calculations for Ar@$C_{60}$ and Xe@$C_{60}$, respectively. The static $\alpha_d^C$ and dynamic $\alpha_d^C(\omega)$ ($\alpha_d^C \equiv \alpha_d^C(0)$) dipole polarizability of $C_{60}$ was taken from previous calculations [4], just as static polarizability of Ar and Xe. Namely, we used the following values $\alpha_d^C = 909$, $\alpha_d^{Ar} = 10.7$ and $\alpha_d^{Xe} = 28$. Having in mind to understand the effects qualitatively, we put $b_C = R = 6.72$, $b_A = r_A$, for Ar $r_{Ar} = 1.98$, and for Xe $r_{Xe} = 2.52$. The thickness of $C_{60}$ is $\Delta = 2.91$. We choose the parameters $U_0$ (assuming $d = \Delta/2$) in (1) in such a way that the electron affinity of $C_{60}^-$ coincide with its experimental value [6].

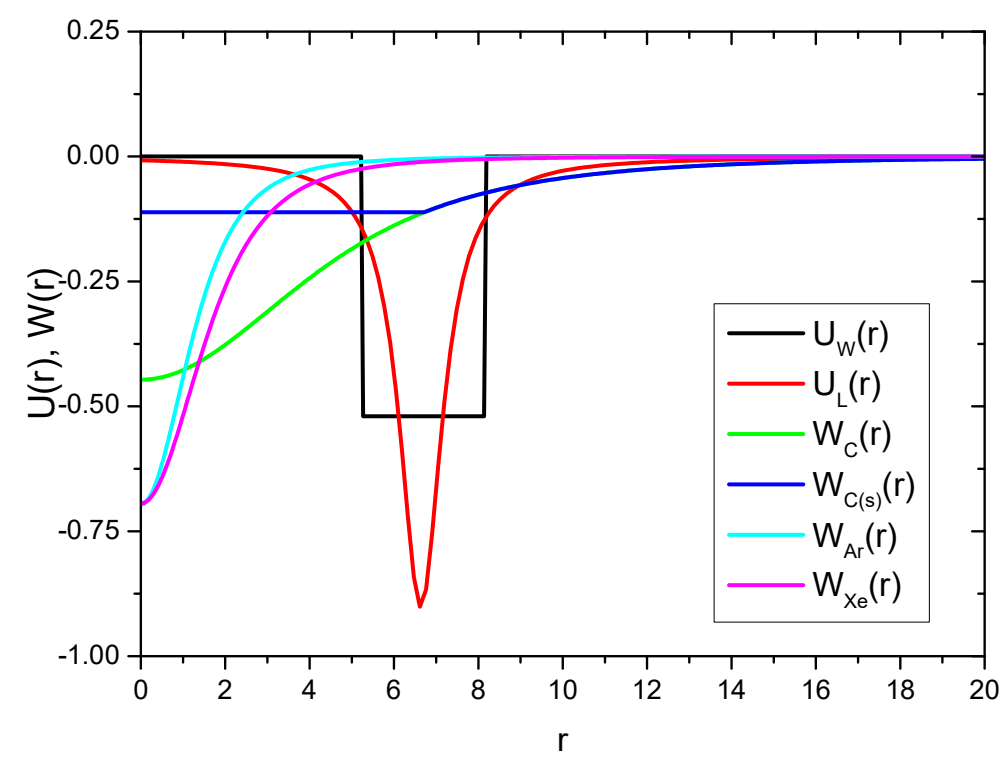

Fig. 1. Potentials, those act upon the atomic electrons in endohedrals.

Fig. 1 demonstrates the potentials engaged in our calculations. Potential $U_W(r)$ presents the $C_{60}$ rectangular well, while $U_L(r)$ presents the Lorentz-shape (1) potential; $W_C(r)$ and $W_{C(s)}(r)$ give the $C_{60}$ polarization potentials, Eq. (2) and (3). Atomic polarization potentials are $W_{Ar}(r)$ and $W_{Xe}(r)$ for Ar and Xe, respectively. We see that $W_{C(s)}(r)$ is considerably smaller than $W_C(r)$ within the region of $r$ that contributes essentially to photoionization cross-section. Let us remind that in our calculations we use $W_{C(s)}(r)$ since it represents $C_{60}$ obviously better than $W_C(r)$. Polarization potentials for Ar and Xe are close to each other.

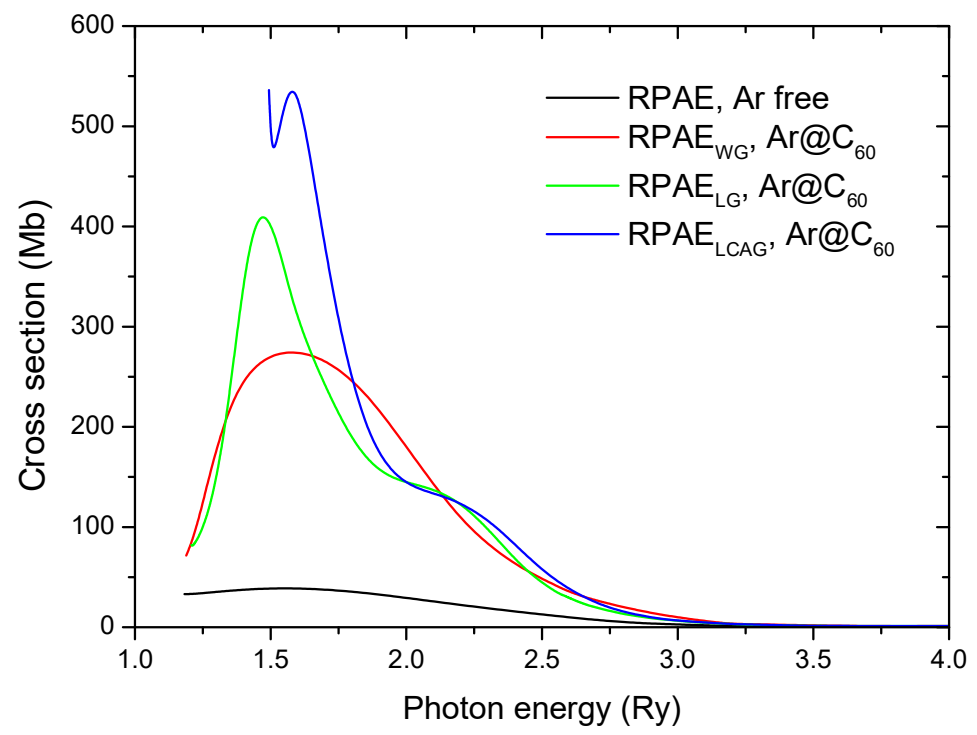

Fig. 2. The RPAE photoionization cross-section of $3p^6$ and $3s^2$ subshells of Ar and Ar@$C_{60}$.

Fig. 2-5 presents the obtained cross-sections. In Fig. 2 we depict the RPAE cross-sections for $3p^6$ and $3s^2$ subshells of Ar and Ar@$C_{60}$. Lower subscripts WG, LG, LCAG denote, respectively, the results that take into account, along with the polarization of $C_{60}$ shell by the photon beam (G) (8), the square well static potential (WG), Lorentz potential (LG) of $C_{60}$, the $C_{60}$ polarization potential (C(s)) (3), and in addition, the atomic potential (A) (4). We denote all that entirely by the subscript LCAG.

We see that the account of $C_{60}$ action increases the cross-sections of A@$C_{60}$ profoundly, as compared to its pure Ar values. Fullerene adds also a prominent oscillation in the cross-section of the endohedral. Note that the inclusion of attractive potentials (1), (3) and (4) shifts the photoionization thresholds to higher energies.

Fig. 3 presents the Xe and Xe@$C_{60}$ photoionization cross-sections for the photon energy region from $3p^6$ threshold to well above the threshold of $4d^{10}$. The notations are the same as in Fig. 2. The presence of $C_{60}$ manifests itself by a tremendous increase of the cross-section near the $3p^6$ threshold and up to 2 Ry above it.

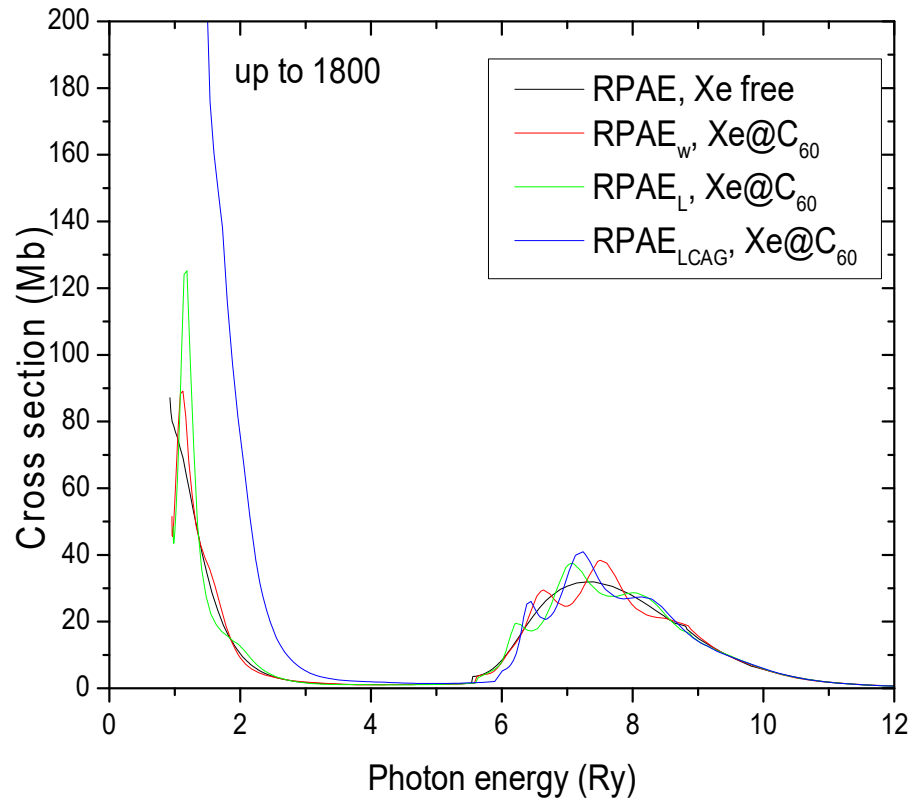

Fig. 3. Photoionization cross-section of Xe and Xe@$C_{60}$ in RPAE.

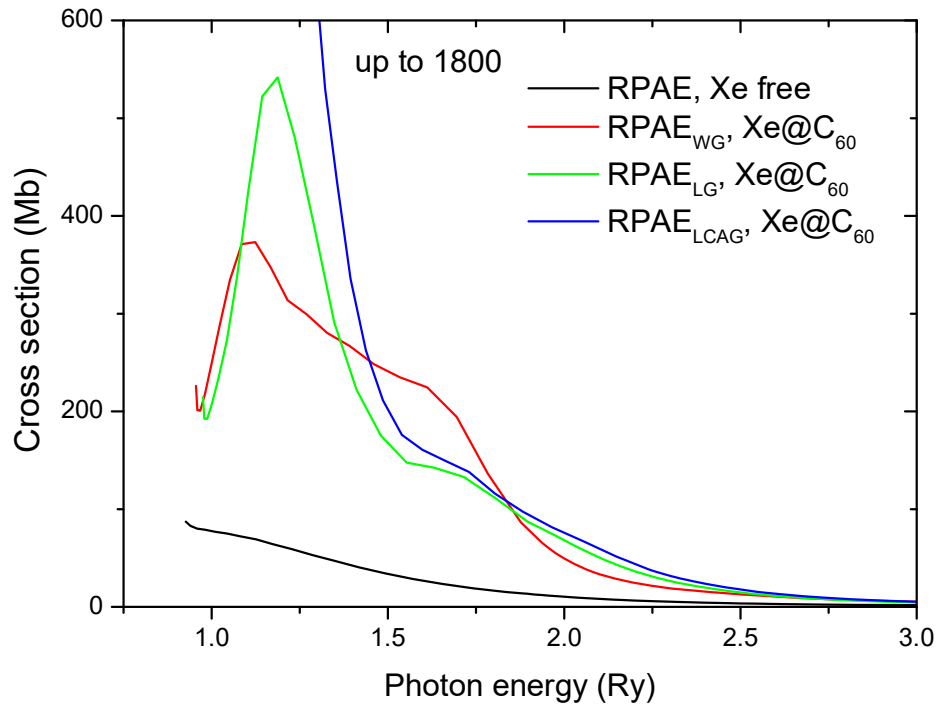

Fig. 4. Photoionization cross-section of the $5p^6$ subshell of Xe and Xe@$C_{60}$ in RPAE.

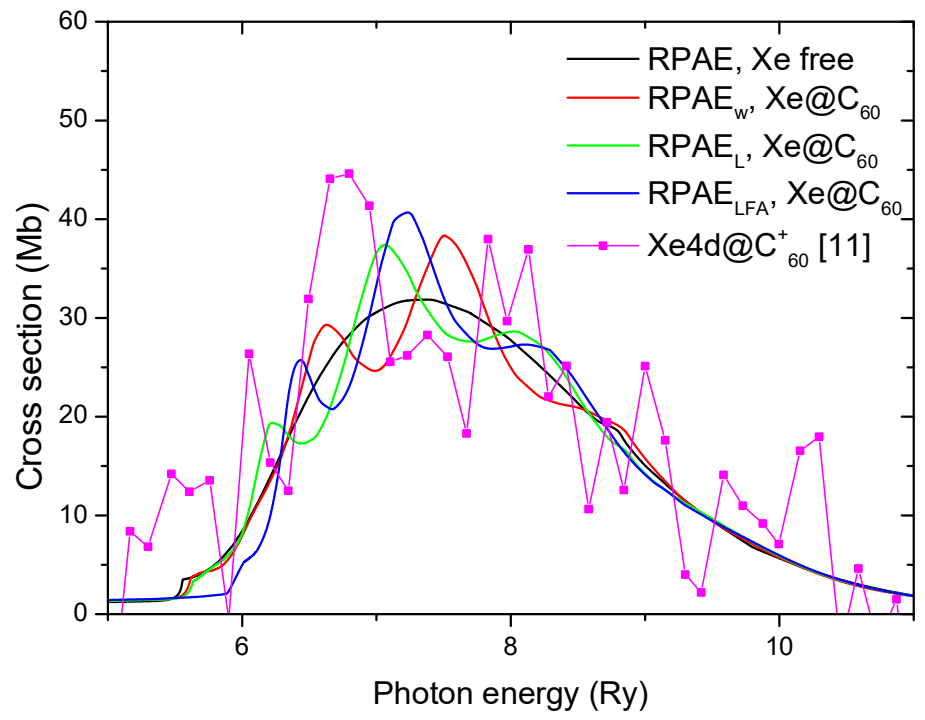

Fig. 5. Photoionization cross-section of the $4d^{10}$ subshell of Xe and Xe@$C_{60}$ in RPAE.

For $4d^{10}$ we see additional oscillations that were predicted some time ago [10] and qualitatively confirmed in recent experiments [11]. The role of $G^C$ factor at $\omega$ above 4Ry is small. As in the case of Ar, the inclusion of attractive potentials (1), (3) and (4) shifts the photoionization thresholds to higher energies.

Fig. 4 presents the photoionization cross-sections of the Xe@$C_{60}$ in the vicinity of the $5p^6$ subshell in more details than Fig. 3. The notations are the same as in previous figures – 2 and 3. We see that the results for Lorentz-type potential are essentially different from that for the square-well potential.

In Fig. 5 we depict the photoionization cross-section of the $4d^{10}$ subshell of Xe and Xe@$C_{60}$ in RPAE. In this photon energy range the factor $G^C$ is close to 1, since the $C_{60}$ shell that consist of 240 collectivized electrons becomes almost totally transparent for photons with energies considerably higher than the position of the $C_{60}$ Giant resonance, i.e. above 2Ry. Fig. 5 includes the only existing experimental data [11].

The main role above the $4d^{10}$ threshold plays the static potential (1). But the polarization potentials (3) and (4) affect noticeably the locations of the additional maximums. However, since due to technical reasons the measurements in [11]. were performed for ions Xe@$C_{60}$+, not for neutrals Xe@$C_{60}$, the theory-experiment comparison is unable to play a decisive role in choosing the best set of parameters in the potential employed.

**4.** In this Letter we present a relatively simple approach that permits to take into account the action of polarization potential of a fullerene $C_{60}$ upon encapsulated atoms Ar and Xe, along with the action of dynamic polarization of $C_{60}$ upon the incoming photon beam.

We suggested a shape of $C_{60}$ polarization potential that fits better to the polarizable object structure than usual polarization potentials, used in studies of electron-endohedral elastic scattering.

We demonstrate previously unnoted overall sufficiently strong, in some cases, profound role of polarization interaction in endohedrals.

It would be very interesting to perform experimental investigation aiming to demonstrate the prominent role of polarization upon photoionization cross-section of the endohedrals. Such experiment would be of great importance for the understanding of electronic structure of endohedrals.

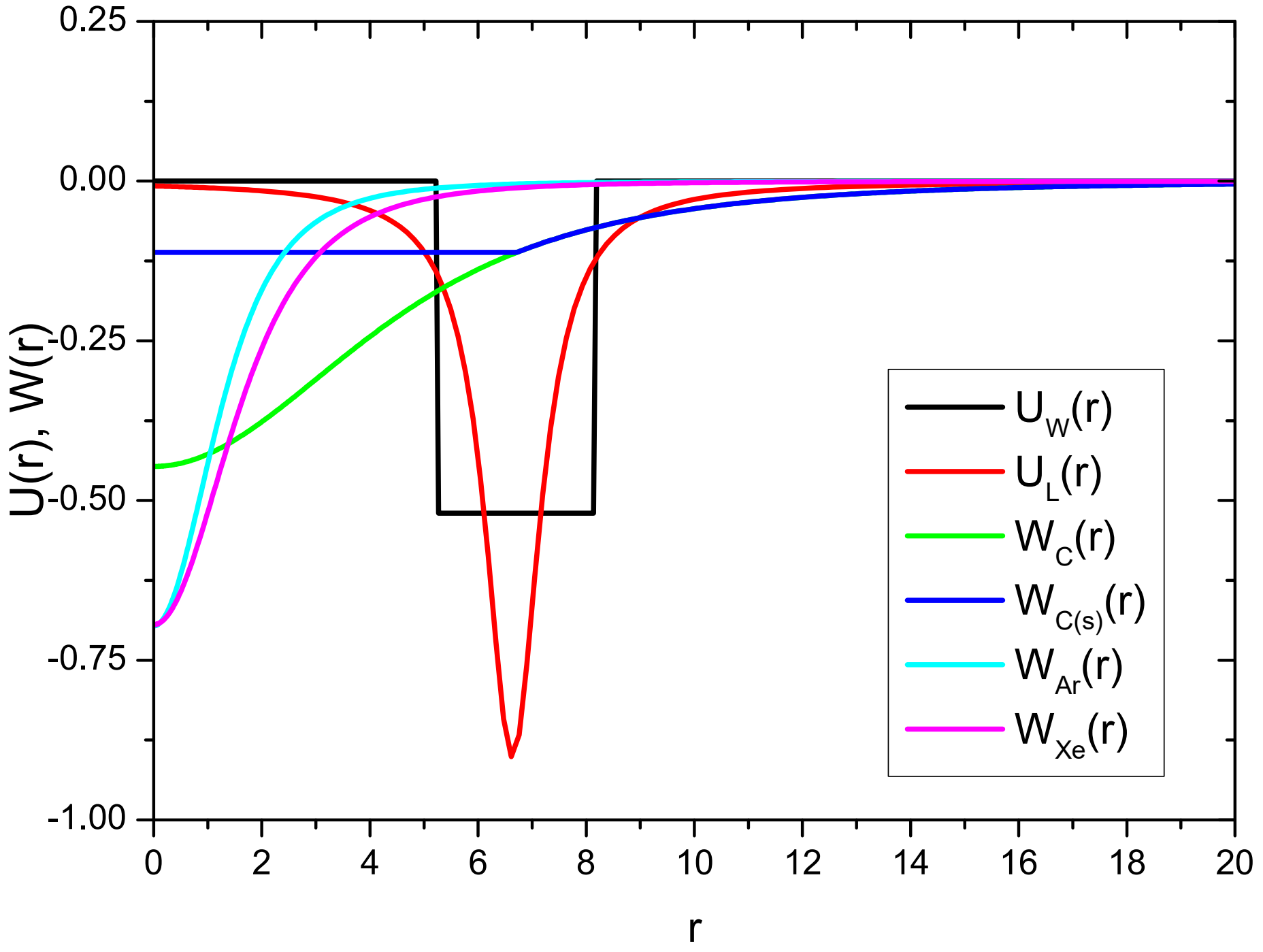

Fig. 1. Potentials, those act upon the atomic electrons in endohedrals.

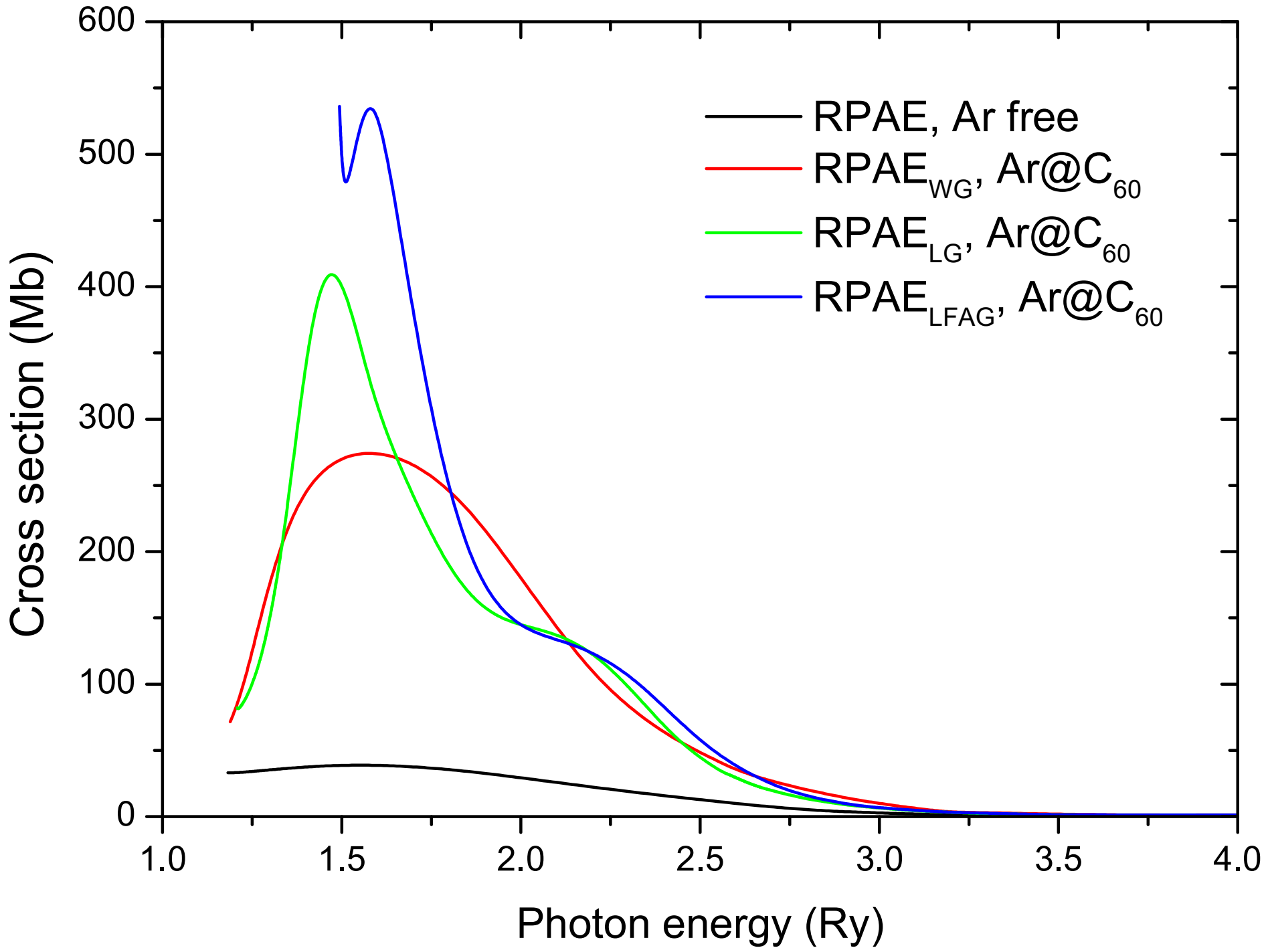

Fig. 2. The RPAE photoionization cross-section of $3p^6$ and $3s^2$ subshell of Ar and Ar@$C_{60}$.

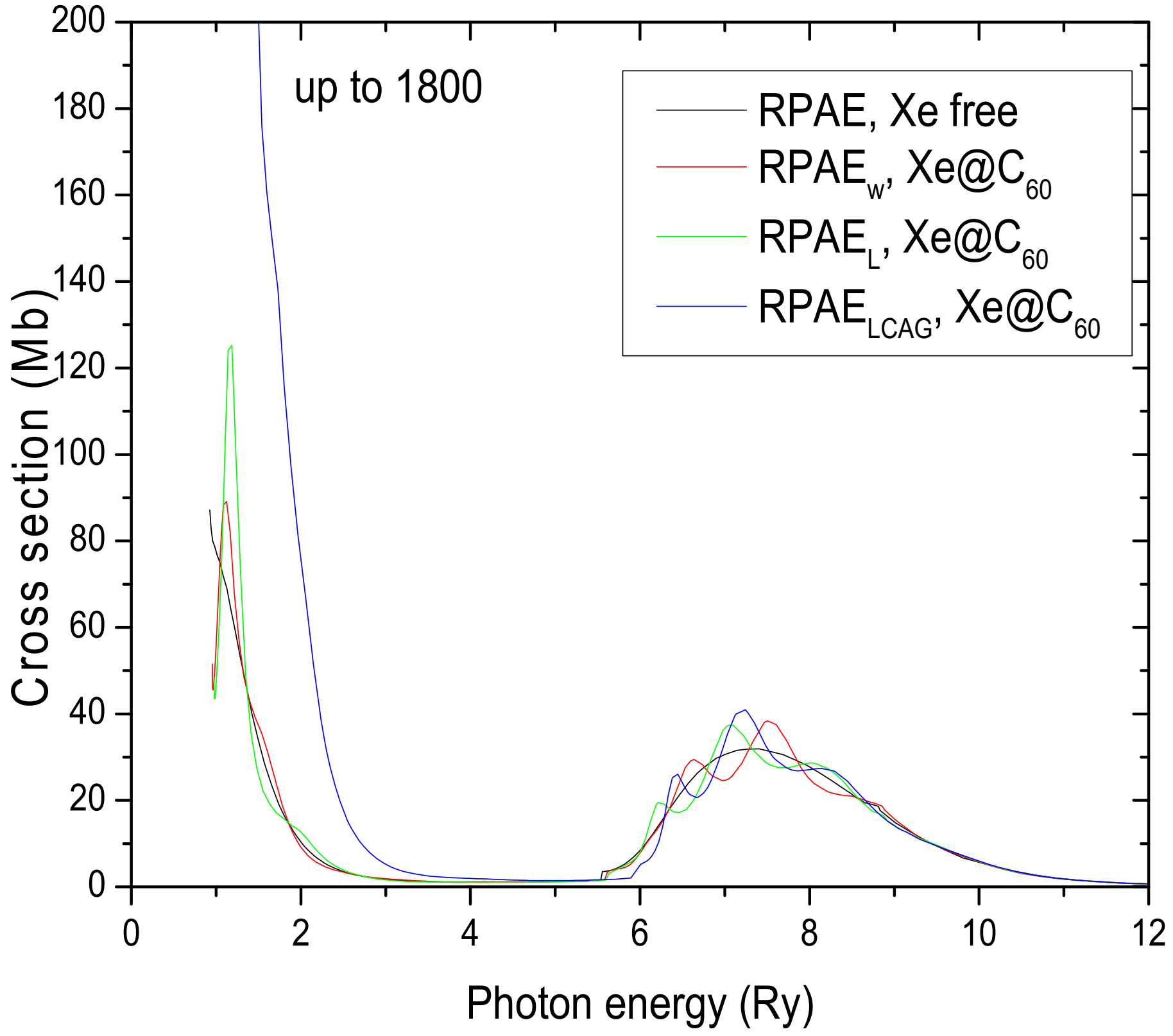

Fig. 3. Photoionization cross-section of Xe and Xe@$C_{60}$ in RPAE.

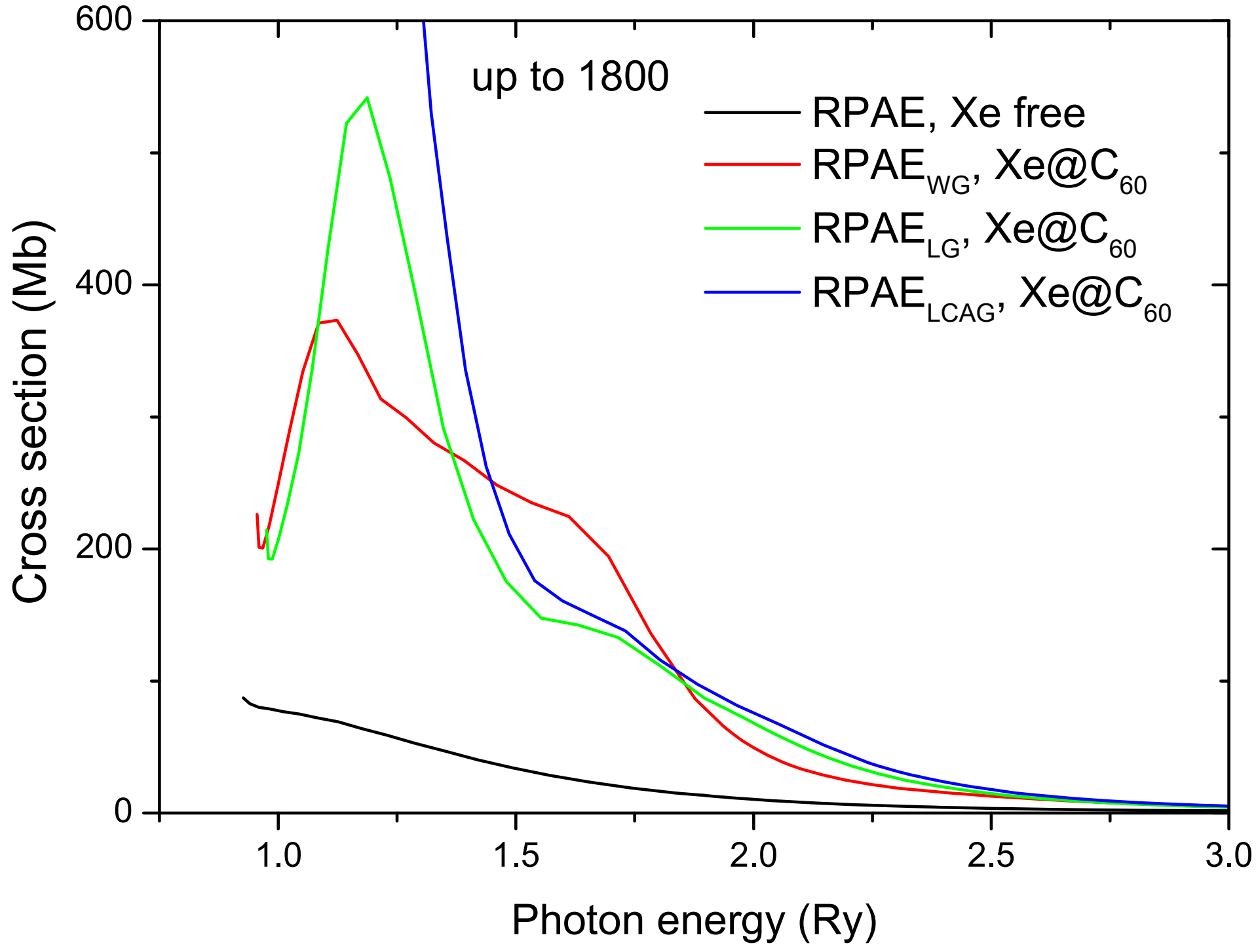

Fig. 4. Photoionization cross-section of the $5p^6$ subshell of Xe and Xe@$C_{60}$.

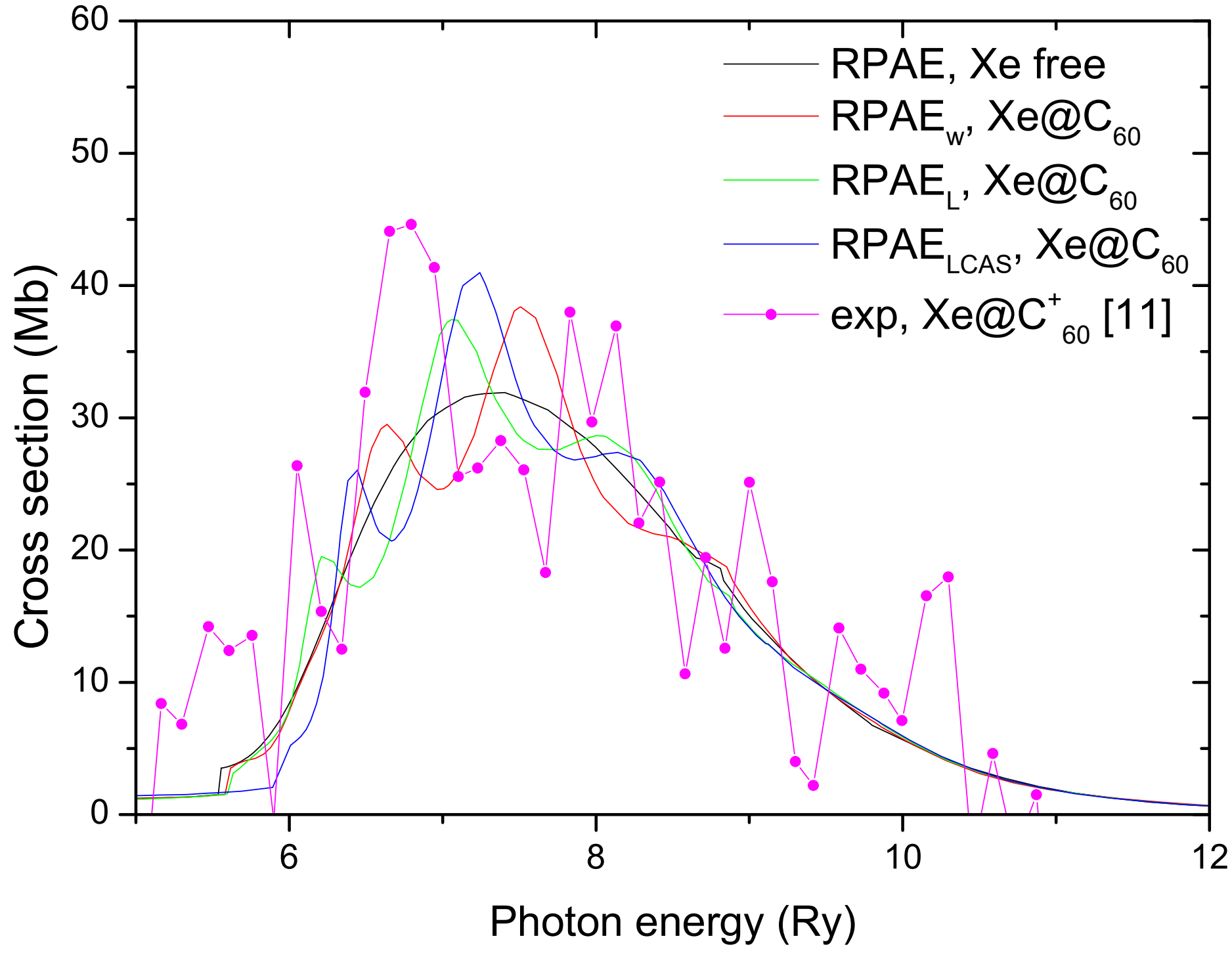

Fig. 5. Photoionization cross-section of the $4d^{10}$ subshell of Xe and Xe@$C_{60}$ in RPAE.